\documentclass[twocolumn,preprintnumbers,amsmath,aps,numbers=noenddot]{revtex4}
\usepackage{graphicx}
\usepackage{dcolumn}
\usepackage{bm}
\usepackage{color}
\begin{document}
\newcommand{\kvec}{\mbox{{\scriptsize {\bf k}}}}
\def\eq#1(\ref{#1)}
\def\fig#1{\ref{#1}}
\preprint{send to: \textcolor{blue}{\bf physica status solidi (b)}}
\title{
---------------------------------------------------------------------------------------------------------------\\
Characterization of the high-pressure superconductivity in the {\it Pnma} phase of calcium}
\author{R. Szcz{\c{e}}{\'s}niak$^{\left(1\right)}$, D. Szcz{\c{e}}{\'s}niak$^{\left(2, 3\right)}$}
\affiliation{1. Institute of Physics, Cz{\c{e}}stochowa University of Technology, Al. Armii Krajowej 19, 42-200 Cz{\c{e}}stochowa, Poland,}
\affiliation{2. Institute of Physics, Jan D{\l}ugosz University in Cz{\c{e}}stochowa, Al. Armii Krajowej 13/15, 42-200 Cz{\c{e}}stochowa, Poland,}
\affiliation{3. Institute for Molecules and Materials UMR 6283, University of Maine, Ave. Olivier Messiaen, 72085 Le Mans, France.}
\email{d.szczesniak@ajd.czest.pl}
%
\date{\today} 
\begin{abstract}
The thermodynamic parameters of the superconducting state in calcium under the pressure at $161$ GPa have been calculated within the framework of the Eliashberg approach. It has been shown that the value of the Coulomb pseudopotential is high ($\mu^{*}_{C}=0.24$) and the critical temperature ($T_{C}=25$ K) should be determined from the modified Allen-Dynes formula. In addition, it has been found that the basic dimensionless ratios of the thermodynamic parameters significantly diverge from the BCS predictions, and take the following values: 
(i) The zero temperature energy gap to the critical temperature ($R_{1}\equiv 2\Delta\left(0\right)/k_{B}T_{C}$) is equal to 4.01. 
(ii) The ratio $R_{2}\equiv \left(C^{S}\left(T_{C}\right)- C^{N}\left(T_{C}\right)\right)/C^{N}\left(T_{C}\right)$ equals $2.17$, where $C^{S}$ and $C^{N}$ denote the specific heats for the superconducting and normal state, respectively. 
(iii) The quantity $R_{3}\equiv T_{C}C^{N}\left(T_{C}\right)/H^{2}_{C}\left(0\right)=0.158$, where $H_{C}$ indicates the thermodynamic critical field. 
Finally, it has been proven that the electron effective mass is large and takes the maximum of $2.32m_{e}$ at $T_{C}$.
\end{abstract}
\maketitle
{\bf Keywords:} Ca-superconductor, High-pressure effects, Thermodynamic properties.

\section{INTRODUCTION}

\begin{figure}[ht]
\centering
\includegraphics[width=\columnwidth]{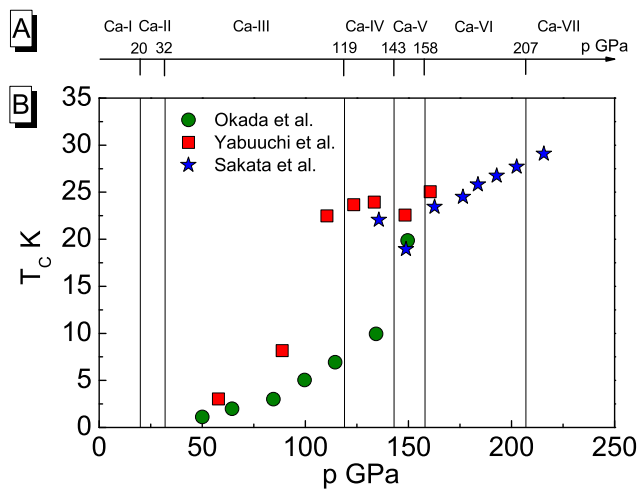}
\caption{(A) The sequence of the structural phase transitions in calcium determined on the basis of the experimental data. (B) The dependence of the critical temperature on the pressure: stars - Okada {\it et al.} \cite{Okada}, squares - Yabuuchi {\it et al.} \cite{Yabuuchi1}, 
circles - Sakata {\it et al.} \cite{Sakata}.}
\label{fig1}
\end{figure}

Under the influence of the high pressure, the elemental calcium undergoes a series of structural phase transitions. In particular, one can distinguish seven phases in the range of pressure ($p$) from $0$ to $241$ GPa \cite{Olijnyk}, \cite{Yabuuchi}, \cite{Sakata} (please see Figure \fig{fig1} (A) for the details).

The two first phases, namely Ca-I and Ca-II, have been classified as a fcc and bcc structures, respectively \cite{Olijnyk}, \cite{Yabuuchi}. The third phase (Ca-III) has been primarily linked with the sc structure, however the recent reports suggest other assignments. On the basis of the theoretical studies, Teweldeberhan {\it et al.} proposed the {\it Cmmm} structure \cite{Teweldeberhan}. Nakamoto {\it et al.} also vote in favor of the {\it Cmmm} structure \cite{Nakamoto}. On the other hand, Mao {\it et al.} have predicted the transition from the sc-like structure to the monoclinic phase at $30$ K and $p\simeq 40$ GPa \cite{Mao}. It needs to be underlined that the stability of the structure sc in the area of the existence of the phase Ca-III is being confirmed by the results achieved by Errea {\it et al.} and Yao {\it et al.}, at least for the temperature of $300$ K \cite{Errea}, \cite{Yao}.

The existence of the phases Ca-IV and Ca-V has been experimentally examined in the papers \cite{Yabuuchi} and \cite{Nakamoto}. Fujihisa {\it et al.} have proposed for them the following assignment: the structure Ca-IV should be characterized by $P4_{1}2_{1}2$ and Ca-V by {\it Cmca} space groups, respectively \cite{Fujihisa}.

In the year $2010$, Nakamoto {\it et al.} discovered the new Ca-VI phase with the {\it Pnma} structure \cite{Nakamoto2}. Further, in the year $2011$, Sakata {\it et al.} have reported the existence of the host-guest phase Ca-VII \cite{Sakata}. We can notice that the high-pressure phase of the host-guest character had been previously predicted by Arapan {\it et al.} and then by Ishikawa {\it et al.} \cite{Arapan}, \cite{Ishikawa}.

The first mention of the existence of the pressure-induced superconducting state in calcium was provided by Dunn and Bundy in 1981 \cite{Dunn}. Fifteen years later, Okada {\it et al.} determined the dependence of the critical temperature ($T_{C}$) on the pressure up to the 150 GPa \cite{Okada} (please see Figure \fig{fig1} (B) for details). In the year 2006, Yabuuchi {\it et al.} have repeated the experimental studies of Okada \cite{Yabuuchi1}. It has been found that the values of the critical temperature increase much faster together with the increase of $p$ in comparison to the results achieved by Okada. The last notable experimental results have been obtained by Sakata {\it et al.} \cite{Sakata}. On the basis of Figure \fig{fig1} (B), it can be easily noticed that for $p=216$ GPa, the critical temperature takes the value equal to $29$ K (the highest observed $T_{C}$ among all elements). 
However, this result has been challenged by Andersson \cite{Andersson}.

In the presented paper, we have determined all relevant thermodynamic parameters of the superconducting state that is induced in calcium under the pressure at $161$ GPa. We draw the readers' attention to the fact that the pressure of $161$ GPa represents the highest value of $p$ considered by Yabuuchi {\it et al.} \cite{Yabuuchi1}. Additionally, the high value of the critical temperature at $p=161$ GPa, which is equal to $\sim 25$ K, has been recently confirmed by the results obtained by Sakata {\it et al.} \cite{Sakata}.   

For the purpose of this paper, we have assumed that the phase Ca-VI is being characterized by the $Pnma$ crystal structure. To support this assumption we quote the results presented in: \cite{Nakamoto2}, \cite{Yin} and \cite{Aftabuzzaman}. 

\section{THE ELIASHBERG EQUATIONS}

On the imaginary axis ($i\equiv\sqrt{-1}$), the order parameter ($\Delta_{n}\equiv\Delta\left(i\omega_{n}\right)$) and the wave function renormalization factor ($Z_{n}\equiv Z\left(i\omega_{n}\right)$) can be calculated by using the Eliashberg equations \cite{Eliashberg}:
\begin{equation}
\label{r1}
\Delta_{n}Z_{n}=\frac{\pi}{\beta} \sum_{m=-M}^{M}
\frac{K\left(\omega_{n}-\omega_{m}\right)-\mu^{*}\theta\left(\omega_{c}-|\omega_{m}|\right)} 
{\sqrt{\omega_m^2+\Delta_m^2}}
\Delta_{m}, 
\end{equation}
\begin{equation}
\label{r2}
Z_n=1+\frac {\pi}{\beta\omega _n }\sum_{m=-M}^{M}
\frac{K\left(\omega_{n}-\omega_{m}\right)}
{\sqrt{\omega_m^2+\Delta_m^2}}\omega_m.
\end{equation}
In equations (\ref{r1}) and (\ref{r2}) the symbol $\omega_{n}\equiv \frac{\pi}{\beta}\left(2n-1\right)$ denotes the $n$-th Matsubara frequency, where $\beta\equiv 1/k_{B}T$, and $k_{B}$ is the Boltzmann constant. 

The complicated form of the electron-phonon pairing kernel is represented by the expression:
\begin{equation}
\label{r3}
K\left(\omega_{n}-\omega_{m}\right)\equiv 2\int_0^{\Omega_{\rm{max}}}d\Omega\frac{\alpha^{2}F\left(\Omega\right)\Omega}
{\left(\omega_n-\omega_m\right)^2+\Omega ^2},
\end{equation}
where $\Omega_{\rm{max}}$ is the maximum phonon frequency ($\Omega_{\rm{max}}=71.37$ meV), and $\alpha^{2}F\left(\Omega\right)$ indicates the Eliashberg function, which models the shape of the electron-phonon interaction in a detailed way. In the presented paper, the form of the $\alpha^{2}F\left(\Omega\right)$ function has been taken from Yin {\it et al.} \cite{Yin}.

The depairing interaction between electrons is described with the use of the Coulomb pseudopotential ($\mu^{*}$). The symbol $\theta$ denotes the Heaviside function and $\omega_{c}$ is the phonon cut-off frequency: $\omega_{c}=3\Omega_{\rm{max}}$.

In the presented paper the Eliashberg equations have been solved for $2201$ Matsubara frequencies ($M=1100$). In this case, the obtained solutions are stable for the temperatures greater than or equal to $T_{0}=5$ K. A detailed discussion of the numerical method has been presented in \cite{Szczesniak1a}-\cite{Szczesniak1e}.

\section{THE COULOMB PSEUDOPOTENTIAL}

The physical value of the Coulomb pseudopotential ($\mu^{*}_{C}$) can be defined by using the condition: $\left[\Delta_{m=1}\right]_{T=T_{C}}=0$, where the critical temperature is equal to the experimental value ($T_{C}=25$ K) \cite{Yabuuchi1}.

\begin{figure}[ht]
\centering
\includegraphics[width=\columnwidth]{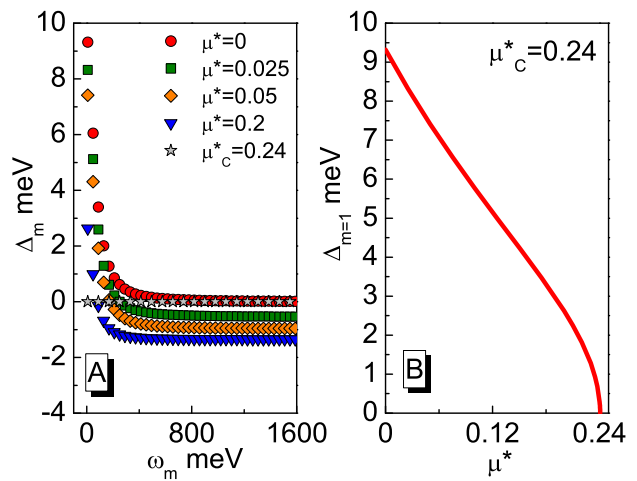}
\caption{(A) The dependence of the order parameter on $\omega_{m}$ for selected values of the Coulomb pseudopotential ($T=T_{C}$). (B) The maximum value of the order parameter as a function of the Coulomb pseudopotential.}
\label{fig2}
\end{figure}

In Figure \fig{fig2} (A) we have presented the dependence of the order parameter on $\omega_{m}$ for selected values of $\mu^{*}$. One can notice that together with the increase of the Coulomb pseudopotential, the largest value of the order parameter ($\Delta_{m=1}$) decreases. Additionally, in Figure \fig{fig2} (B) we have outlined the complete form of the function $\Delta_{m=1}\left(\mu^{*}\right)$. On the basis of the obtained results, we have found that the physical value of the Coulomb pseudopotential is equal to $0.24$.

The above result means that the depairing electron correlations in calcium are relatively strong (the classical low-temperature superconductors $\mu^{*}_{C}$ is about $0.1$ \cite{Carbotte}). It can be noted that similar non-standard value of $\mu^{*}_{C}$  has been obtained for lithium and $\rm CaLi_{2}$ \cite{Jishi}, \cite{Profeta}, \cite{Luders}, \cite{Szczesniak1f}, \cite{Szczesniak1g}. For example, the properties of the superconducting state in the fcc phase of lithium for the pressure values $22.3$ GPa ($T_{C}= 7.27$ K) and $29.7$ GPa ($T_{C}=13.93$ K) have been specified in the paper \cite{Szczesniak1f}. It has been shown that the physical value of the Coulomb pseudopotential increases with $p$ from $0.22$ to $0.36$. In the case of $\rm CaLi_{2}$, the parameter $\mu^{*}_{C}$ is equal to $0.23$ ($p=45$ GPa and $T_{C}=12.9$ K) \cite{Szczesniak1g}.

The high values of $\mu^{*}_{C}$ in Ca, Li, and $\rm CaLi_{2}$ are difficult to explain in the framework of the classical Morel-Anderson (MA) model  \cite{Morel}, where: $\mu^{*}_{C}=\mu\left[1+\mu\ln\left(\omega_{e}/\omega_{ph}\right)\right]^{-1}$. The symbol $\mu$ is defined by: $\mu\equiv\rho\left(0\right)U_{C}$, where $\rho\left(0\right)$ is the electronic density of states at the Fermi level and $U_{C}$ is the Coulomb potential; $\omega_{e}$ and $\omega_{ph}$ denote the characteristic electron and phonon frequency, respectively. Since $\omega_{e}\gg\omega_{ph}$, the MA pseudopotential is of the order $0.1$ and $\mu^{*}_{C}\ll\mu$. 

It can be noted that the MA model corresponds to treating the irreducible vertex to the first order in $U_{C}$. Recently, Bauer, Han, and Gunnarsson have extended the MA theory to the second order in $U_{C}$. The main result is that the retardation effects lead to the reduction $\mu\rightarrow\mu^{*}_{C}$ also in the higher order calculation, but not as efficiently as in the first order \cite{Bauer}. The model presented in the paper \cite{Bauer} is probably the most advanced attempt to explain the high values of $\mu^{*}_{C}$ in Ca, Li, and $\rm CaLi_{2}$.

In particular, Bauer {\it et al.} have given the following expression for the physical value of the Coulomb pseudopotential:
\begin{equation}
\label{rDodatkowe}
\mu^{*}_{C}=\frac{\mu+a\mu^{2}}{1+\mu\ln\left[\frac{\omega_{e}}{\omega_{ph}}\right]+a\mu^{2}\ln\left[\frac{\alpha\omega_{e}}{\omega_{ph}}\right]},
\end{equation}
where $a=1.38$ and $\alpha\simeq 0.10$.

On the basis of the equation (\ref{rDodatkowe}) one can easily estimate the value of $U_{C}$ for the real materials. In the paper, we assume the following: $\omega_{e}=W$ ($W$ is the half-band width), $\omega_{ph}=\omega_{{\rm ln}}$ ($\omega_{{\rm ln}}\equiv \exp\left[\frac{2}{\lambda}\int^{\Omega_{\rm{max}}}_{0}d\Omega\frac{\alpha^{2}F\left(\Omega\right)}
{\Omega}\ln\left(\Omega\right)\right]$), and $\rho\left(0\right)=1/2W$ (the constant DOS). In the literature, the values of all the important  parameters are provided only for ${\rm CaLi_{2}}$ ($p=45$ GPa). In particular: $\mu_{C}^{*}=0.23$, $W=1991$ meV, and $\omega_{\rm ln}=17.02$ meV  \cite{Szczesniak1g}, \cite{Xie}. The result is the following: $U_{C}=2803$ meV.  

Next, we address an important issue, namely, how big is the error bar of the calculated physical value of the Coulomb pseudopotential ($\Delta\mu^{*}_{C}$). First of all, we can notice that in the framework of the presented analysis, the value of  $\mu^{*}_{C}$ depends on the shape of the Eliashberg function and the accuracy of the experimental value of $T_{C}$. For calcium the appropriate Eliashberg function, taken from \cite{Yin}, has been calculated by using the linear-response method (full-potential LMTART code \cite{Savrasov}).  In that paper, the Eliashberg function error bar has been omitted ($\left[\Delta\mu^{*}_{C}\right]_{\alpha^{2}F\left(\Omega\right)}=0$). On the other hand, the value of the critical temperature has been measured with the accuracy about $\pm 1$ K  \cite{Yabuuchi1}. On the basis of these facts, we have obtained: 
$\left[\Delta\mu^{*}_{C}\right]_{T_{C}}=\pm 0.02$.

\section{THE CRITICAL TEMPERATURE}

%
\begin{figure}[ht]
\centering
\includegraphics[width=\columnwidth]{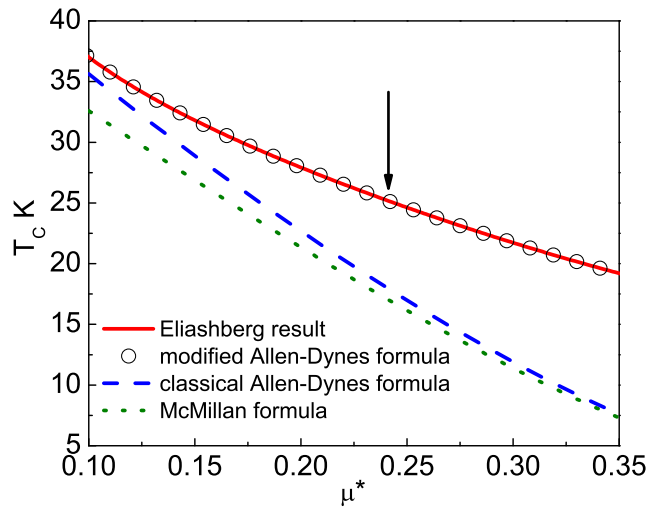}
\caption{The dependence of the critical temperature on the Coulomb pseudopotential. The filled circles mean the results obtained by using the Eliashberg equations; the arrow indicates the experimental value of the critical temperature ($\mu^{*}_{C}=0.24$). The solid line denotes the results obtained with the help of the modified Allen-Dynes formula. Finally, the dashed and dotted lines have been generated based on the classical Allen-Dynes formula and the McMillan expression, respectively.}
\label{fig3}
\end{figure}

In the framework of the presented formalism, the exact value of the critical temperature should be obtained on the basis of Eliashberg equations. However, in the case of the data interpretation, it is far more convenient to use the simple formula that explicitly reproduces the results of the advanced numerical calculations. In the branch literature there are known two basic formulas that serve for the determination of the critical temperature's value. The first one has been introduced by McMillan \cite{McMillan}; the second one is the Allen-Dynes expression \cite{AllenDynes}. Unfortunately, in the case of calcium, both formulas considerably underestimate the critical temperature. According to the above, the Allen-Dynes formula has been modified in such a way, that it allows us to reproduce the numerical results correctly. 
Particularly, in order to achieve the proper values of the fitting parameters, the dependence of the critical temperature on the Coulomb pseudopotential has been analyzed on the level of the Eliashberg equations (only $\alpha^{2}F\left(\Omega\right)$ has been considered as the physical input parameter). Next, the least-squares method was applied. The obtained result is presented below:

\begin{equation}
\label{r4}
k_{B}T_{C}=f_{1}\left(\mu^{*}\right)f_{2}\left(\mu^{*}\right)\frac{\omega_{\rm ln}}{1.45}\exp\left[\frac{-1.03\left(1+\lambda\right)}{\lambda-\mu^{*}\left(1+0.06\lambda\right)}\right],
\end{equation}
where the functions $f_{1}\left(\mu^{*}\right)$ and $f_{2}\left(\mu^{*}\right)$ are expressed by the formulas:
\begin{equation}
\label{r5a}
f_{1}\left(\mu^{*}\right)\equiv\left[1+\left(\frac{\lambda}{\Lambda_{1}\left(\mu^{*}\right)}\right)^{\frac{3}{2}}\right]^{\frac{1}{3}},
\end{equation}
and
\begin{equation}
\label{r5b}
f_{2}\left(\mu^{*}\right)\equiv 1+\frac{\left(\frac{\sqrt{\omega_{2}}}{\omega_{\rm{ln}}}-1\right)\lambda^{2}}{\lambda^{2}+\Lambda^{2}_{2}\left(\mu^{*}\right)}.
\end{equation}
The parameters, that depend on the Eliashberg function, can be determined on the basis of the expressions:   $\omega_{2}\equiv\frac{2}{\lambda}\int^{\Omega_{\rm{max}}}_{0}d\Omega\alpha^{2}F\left(\Omega\right)\Omega$ and $\lambda\equiv 2\int^{\Omega_{\rm{max}}}_{0}d\Omega\frac{\alpha^{2}F\left(\Omega\right)}{\Omega}$. 
For calcium under the pressure at $161$ GPa, we have respectively: $\sqrt{\omega_{2}}=34.36$ $\rm{meV}$ and $\lambda=1.27$. The fitting functions $\Lambda_{1}\left(\mu^{*}\right)$ and $\Lambda_{2}\left(\mu^{*}\right)$ are presented in the following forms:
\begin{equation}
\label{r6}
\Lambda_{1}\left(\mu^{*}\right)\equiv 0.145(1+115.862\mu^{*}),
\end{equation}
and
\begin{equation}
\label{r7}
\Lambda_{2}\left(\mu^{*}\right)\equiv 5.185(1-2.247\mu^{*})\left(\frac{\sqrt{\omega_2}}{\omega_{\ln}}\right).
\end{equation}

In Figure \fig{fig3} we have presented the numerical solutions obtained with the use of the Eliashberg equations and the modified Allen-Dynes formula. Additionally, for the comparison purposes, we have depicted the results based on the classical formulas derived by Allen-Dynes and McMillan. On the basis of Figure \fig{fig3}, one can observe, that the expression (4) perfectly reproduces the exact Eliashberg numerical predictions.

The constants in the equation (4) deviate notably from the original parameterization. This situation is connected with the fact that the analysis based on the real-axis Eliashberg equations suggests only the semiphenomenological form of the $T_{C}$-formula: 
$k_{B}T_{C}=\frac{\omega_{\rm ln}}{a}\exp\left[-\frac{-b\left(1+\lambda\right)}{\lambda-\mu^{\star}\left(1+c\lambda\right)}\right]$ (see the detailed discussion in \cite{Carbotte}, p. 1051-1052). In the case of calcium under the pressure at $161$ GPa, the values of the Allen-Dynes parameters are inappropriate. Thus, the constants ($a\sim c$) should be fit to the data taken from the exact solutions of the Eliashberg equations. We can notice that the change of the coefficient in the expression (4) under $\omega_{\ln}$ from $1.2$ to $1.45$ slightly lowers the phonon frequency; the two remaining parameters (1.03 and 0.06), in comparison to the classical parameterization, increase the value of the effective electron-phonon coupling constant.  

Moreover, the parameterization of the strong-coupling correction function ($f_{1}\left(\mu^{*}\right)$) and the shape correction function ($f_{2}\left(\mu^{*}\right)$) also deviates from the original form. The achieved result indicates that for the high-pressure superconducting state in calcium the shape function has greater significance than in the classical superconductors. 

The value of the critical temperature for $p=160$ GPa has been also calculated in the paper \cite{Yin}. By using the Allen-Dynes formula the Authors have qualitatively reconstructed the experimental value of $T_{C}$. However, in the examined case the physical value of the Coulomb pseudopotential has been strongly lowered ($\mu^{*}_{C}\sim 0.15$). 

In the last step, we boldly underline that in the literature exist other calculations of $\lambda$ than has been presented in the paper \cite{Yin}. In particular, Lei {\it et al.} have suggested a very large value of the electron-phonon coupling constant ($\lambda=3.75$ for $p=155$ GPa and the sc structure) \cite{Lei}. On the other hand, Aftabuzzaman and Islam have predicted $\lambda=0.903$ for $p=161$ GPa  and the {\it Pnma} structure \cite{Aftabuzzaman}. The last result is similar to the result obtained by Yin {\it et al.} \cite{Yin}.

\section{THE CHARACTERISTICS OF THE SOLUTIONS ON THE IMAGINARY AXIS}

\begin{figure}[ht]
\centering
\includegraphics[width=\columnwidth]{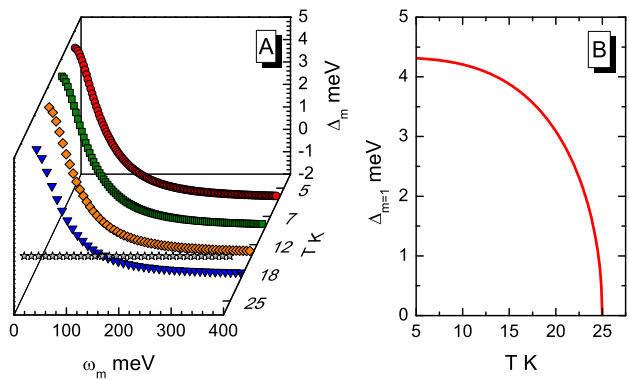}
\caption{(A) The dependence of the order parameter on $\omega_{m}$ for selected values of the temperature. (B) The maximum value of the order parameter as a function of the temperature.}
\label{fig4}
\end{figure}

The form of the order parameter on the imaginary axis for selected values of the temperature has been presented in Figure \fig{fig4} (A). It has been shown that together with the increase of $\omega_{m}$ the absolute values of $\Delta_{m}$ are decreasing and are subjected to saturation. It should be underlined that taking the negative values by the order parameter function is connected with the non-zero value of the Coulomb pseudopotential.

When analyzing the temperature's dependence of the order parameter, we found that absolute values of the function $\Delta_{m}$ decrease together with the temperature's growth. The above result means that together with the growth of the temperature, the less number of Matsubara frequencies give significant contribution to the Eliashberg equations.

The full dependence of the maximum value of the order parameter ($\Delta_{m=1}$) on the temperature has been plotted in Figure \fig{fig4} (B). We can observe that the values $2\Delta_{m=1}\left(T\right)$ with a good approximation reproduce the temperature dependence of the energy gap at the Fermi level.

\begin{figure}[ht]
\centering
\includegraphics[width=\columnwidth]{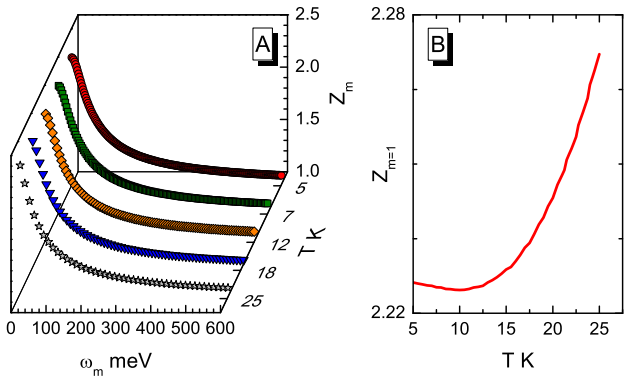}
\caption{(A) The dependence of the wave function renormalization factor on $\omega_{m}$ for selected values of the temperature. (B) The maximum value of the wave function renormalization factor as a function of the temperature.}
\label{fig5}
\end{figure}

In Figure \fig{fig5} (A) we have presented the form of the wave function renormalization factor on the imaginary axis. Similarly as for the order parameter, the increase of $\omega_{m}$ causes the decrease of the successive values of the function $Z_{m}$. In the case of the high values of $\omega_{m}$, the function $Z_{m}$ is subjected to the saturation and takes the value equal to one.

Further, Figure \fig{fig5} (B) presents the full dependence of the maximum value of the wave function renormalization factor on the temperature. It can be noted that the presented function with a good approximation determines the temperature dependence of the electron effective mass. Moreover, from the obtained results, we can conclude that the electron effective mass takes a high value in the entire area of the existence of the superconducting state.

%

\section{THE PHYSICAL VALUE OF THE ORDER PARAMETER}

In order to determine the physical value of the order parameter for the chosen temperature, the solutions of the Eliashberg equations on the imaginary axis ($i\omega_{n}$) should be analytically continued on the real axis ($\omega$). In the presented paper we have used the method introduced by Beach {\it et al.} \cite{Beach}. The form of the order parameter on the real axis is being reproduced by using the function: 
\begin{equation}
\label{r8}
\Delta\left(\omega\right)=\frac{p_{\Delta 1}+p_{\Delta 2}\omega+...+p_{\Delta r}\omega^{r-1}}
{q_{\Delta 1}+q_{\Delta 2}\omega+...+q_{\Delta r}\omega^{r-1}+\omega^{r}},
\end{equation}
where $p_{\Delta j}$ and $q_{\Delta j}$ denote the number coefficients, and $r=550$.
 
\begin{figure}[ht]
\centering
\includegraphics[width=\columnwidth]{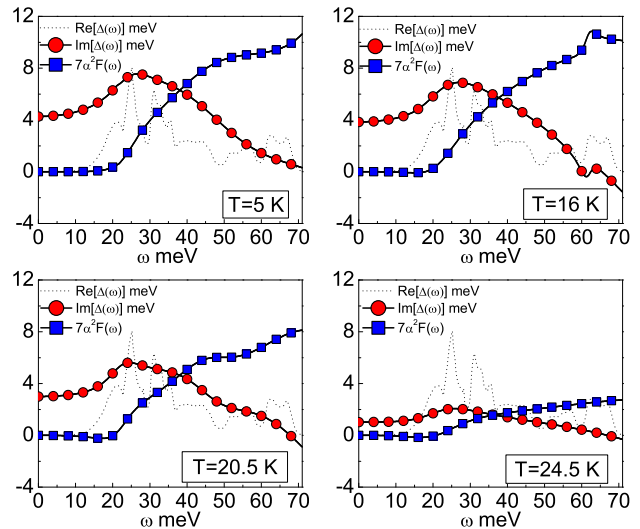}
\caption{The real and imaginary part of the order parameter on the real axis for selected values of the temperature. The rescaled Eliashberg function has been also presented.}
\label{fig6}
\end{figure}

The dependence of the real and imaginary part of the order parameter on the frequency for selected values of the temperature has been presented in Figure \fig{fig6}. Additionally, the rescaled Eliashberg function has been specified. On the basis of the presented results, one can observe that in the range of the low frequencies (from $0$ to about $20$ meV), only the real part of the order parameter takes the non-zero values. From the physical point of view, the obtained result defines the lack of the damping effects. For the higher values of the frequency (from about $20$ meV to about $40$ meV), the real part of the order parameter takes relatively high values, which are clearly induced by the characteristic peaks in the Eliashberg function. Furthermore, we can notice that in the discussed range of the energy, the imaginary part of the order parameter becomes non-zero and strongly increases together with the increase of frequency. For the higher frequencies (above $40$ meV), the real part of the order parameter begins to vanish. This fact is related to the extinction of the Eliashberg function itself.

The physical value of the order parameter for the chosen temperature should be determined on the basis of the expression \cite{Eliashberg}, \cite{Carbotte}:

\begin{equation}
\label{r9}
\Delta\left(T\right)={\rm Re}\left[\Delta\left(\omega=\Delta\left(T\right),T\right)\right].
\end{equation}

In the case of the superconductors the most interesting is the value of the order parameter for the temperature of zero Kelvin ($\Delta\left(0\right)\simeq\Delta\left(T_{0}\right)$). On the basis of the simple calculations we have made the following estimation: $\Delta\left(0\right)=4.32$ meV. 

Let us mention that familiarity with the value of $\Delta\left(0\right)$ and $T_{C}$ allows to calculate the dimensionless ratio: $R_{1}\equiv 2\Delta\left(0\right)/k_{B}T_{C}$. In the case of calcium we have obtained the following: 
\begin{equation}
\label{r10}
R_{1}=4.01.
\end{equation}
The above result indicates that $R_{1}$ considerably exceeds the value predicted by the BCS theory: $\left[R_{1}\right]_{\rm BCS}=3.53$ \cite{BCS}. 

\section{THE ELECTRON EFFECTIVE MASS}

The influence of the electron-phonon interaction on the electron effective mass ($m^{*}_{e}$) can be determined on the basis of the expression: $m^{*}_{e}={\rm Re}\left[Z\left(0\right)\right]m_{e}$, where the symbol $Z\left(0\right)$ denotes the value of the wave function renormalization factor on the real axis and $m_{e}$ is the bare electron mass. 
The form of the wave function renormalization factor on the real axis has been calculated with the use of the analytical continuation method:
\begin{equation}
\label{r11}
Z\left(\omega\right)=\frac{p_{Z 1}+p_{Z 2}\omega+...+p_{Z r}\omega^{r-1}}
{q_{Z 1}+q_{Z 2}\omega+...+q_{Z r}\omega^{r-1}+\omega^{r}},
\end{equation}
where $p_{Z j}$ and $q_{Z j}$ are the number coefficients, and $r=550$.

\begin{figure}[ht]
\centering
\includegraphics[width=\columnwidth]{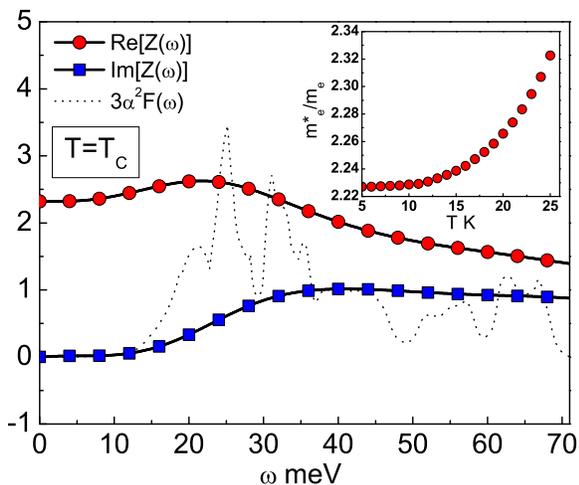}
\caption{The real and imaginary part of the wave function renormalization factor on the real axis. Additionally, the rescaled Eliashberg function has been outlined. The inset represents the dependence of $m^{*}_{e}/m_{e}$ on the temperature.}
\label{fig7}
\end{figure}

In Figure \fig{fig7} we have presented the shape of the function Re$\left[Z\left(\omega\right)\right]$ and Im$\left[Z\left(\omega\right)\right]$ for the critical temperature. Similarly to the situation which took place in the case of the order parameter, for the low frequencies the non-zero is only the real part of the wave function renormalization factor. In the energy range around $20$ meV we can observe characteristic but not so strong amplification of Re$\left[Z\left(\omega\right)\right]$, which is clearly correlated with the peaks of the Eliashberg function. Additionally, the function Im$\left[Z\left(\omega\right)\right]$ is non-zero. In the range of the high frequencies, Re$\left[Z\left(\omega\right)\right]$ decreases together with the increase of $\omega$. 

Next, the dependence of the ratio $m^{*}_{e}/m_{e}$ on the temperature has been determined. The results have been presented in the inset in Figure \fig{fig7}. We have found that the electron effective mass is large in the entire range, in which the superconducting state exists, and reaches its maximum equal to $2.32$ for $T=T_{C}$.

We can notice that for $T=T_{C}$ the value of the ratio $m^{*}_{e}/m_{e}$ can be calculated with a great accuracy by using the simple formula: $m^{*}_{e}/m_{e}\simeq 1+\lambda=2.27$. The consistency between the exact numerical result and the analytical approach is the measure of the presented analysis.

From the physical point of view, the result presented above is particularly important, since it can be verified in a simple way if the measurement of the Sommerfeld coefficient is to be made in the future.
   
\section{THE THERMODYNAMIC CRITICAL FIELD AND THE SPECIFIC HEAT}

The thermodynamic critical field ($H_{C}$) and the difference between the specific heat in the superconducting and normal state 
($\Delta C\equiv C^{S}-C^{N}$) can be calculated on the basis of the free energy difference ($\Delta F\equiv F^{S}-F^{N}$):
\begin{eqnarray}
\label{r12}
\frac{\Delta F}{\rho\left(0\right)}&=&-\frac{2\pi}{\beta}\sum_{m=1}^{M}
\left(\sqrt{\omega^{2}_{m}+\Delta^{2}_{m}}- \left|\omega_{m}\right|\right)\\ \nonumber
&\times&(Z^{{\rm S}}_{m}-Z^{N}_{m}\frac{\left|\omega_{m}\right|}
{\sqrt{\omega^{2}_{m}+\Delta^{2}_{m}}}).  
\end{eqnarray}  

\begin{figure}[ht]
\centering
\includegraphics[width=\columnwidth]{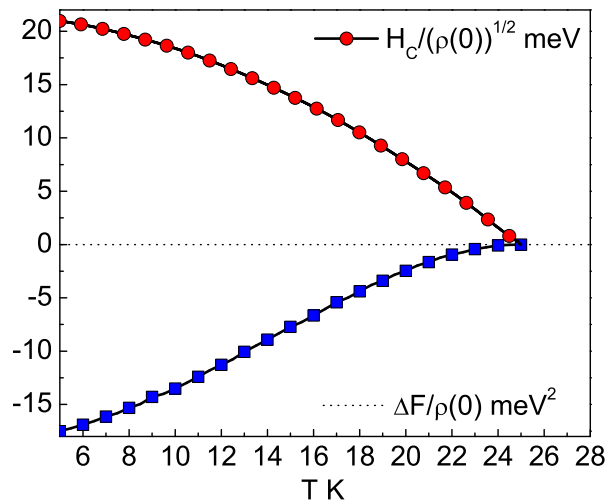}
\caption{The ratios $\Delta F/\rho\left(0\right)$ and $H_{C}/\sqrt{\rho\left(0\right)}$ as a function of the temperature.}
\label{fig08}
\end{figure}

The dependence of the free energy difference on the temperature has been presented in Figure \fig{fig08}. We can see that in the whole range of the existence of the superconducting phase, the value of the ratio $\Delta F/\rho\left(0\right)$ is negative. From the physical point of view, it means that the superconducting state is thermodynamically stable.

The thermodynamic critical field should be determined on the basis of the expression:  
\begin{equation}  
\label{r13}
\frac{H_{C}}{\sqrt{\rho\left(0\right)}}=
\sqrt{-8\pi\left[\Delta F/\rho\left(0\right)\right]}.
\end{equation}
The influence of the temperature on the value of the ratio $H_{C}/\sqrt{\rho\left(0\right)}$ has been presented in the Figure \fig{fig08}.

The difference of the specific heat has been determined on the basis of the formula:
\begin{equation}
\label{r14}
\frac{\Delta C}{k_{B}\rho\left(0\right)}
=-\frac{1}{\beta}\frac{d^{2}\left[\Delta F/\rho\left(0\right)\right]}
{d\left(k_{B}T\right)^{2}}.
\end{equation}
Additionally, the values of the specific heat in the normal state have been also determined:
\begin{equation}
\label{r15}
\frac{C^{N}}{ k_{B}\rho\left(0\right)}=\frac{\gamma}{\beta}, 
\end{equation}
where $\gamma\equiv\frac{2}{3}\pi^{2}\left(1+\lambda\right)$.

\begin{figure}[ht]
\centering
\includegraphics[width=\columnwidth]{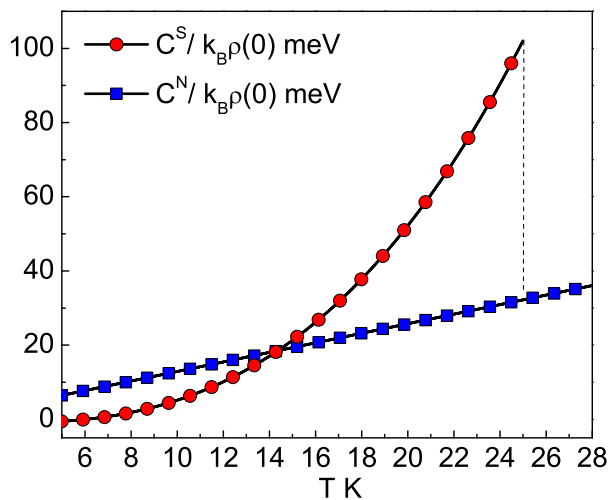}
\caption{The specific heat in the superconducting state and in the normal state as a function of the temperature.}
\label{fig09}
\end{figure}

In Figure \ref{fig09}, we have plotted the dependence of the specific heat in the superconducting state and the normal state on the temperature. The characteristic "jump", which appears at the critical temperature, can be easily noticed.

On the basis of the specified thermodynamic functions, we have calculated the values of the dimensionless ratios: $R_{2}\equiv \Delta C\left(T_{C}\right)/C^{N}\left(T_{C}\right)$ and $R_{3}\equiv T_{C}C^{N}\left(T_{C}\right)/H^{2}_{C}\left(0\right)$. We have obtained the following:
\begin{equation}
\label{r16}
R_{2}=2.17, 
\end{equation}
and
\begin{equation}
\label{r17}
R_{3}=0.158. 
\end{equation}

Taking into account the results above, we can state that the values of the considered ratios significantly diverge from the values predicted by the classical BCS theory. In particular: $\left[R_2\right]_{\rm BCS}=1.43$ and $\left[R_3\right]_{\rm BCS}=0.168$.

\section{SUMMARY}

In the paper, we have determined all relevant thermodynamic parameters of the superconducting state in calcium under the pressure at $161$ GPa.
We have conducted all numerical calculations in the framework of the Eliashberg formalism, where the electron-phonon spectral function $\alpha^{2}F\left(\Omega\right)$ has been taken form the paper \cite{Yin}. On the basis of the exact numerical results, we can state that the depairing electron correlations in calcium are relatively strong ($\mu^{*}_{C}=0.24$). In the next step, the values of the parameters in the Allen-Dynes formula have been calculated. It has been shown that the critical temperature is properly determined by the modified Allen-Dynes expression.

Furthermore, we have proven that the thermodynamic properties of the superconducting state significantly diverge from the predictions based on the simple BCS theory. In particular, the following values of the thermodynamic ratios have been obtained: $R_{1}=4.01$, $R_{2}=2.17$, and $R_{3}=0.158$.
 
In the last step, we have shown that the electron effective mass is large in the entire area of the existence of the superconducting state, and $\left[m^{*}_{e}\right]_{\rm max}=2.32m_{e}$ at $T=T_{C}$.

\begin{acknowledgments}
The authors would like to thank Professor K. Dzili{\'{n}}ski, Professor Z. B{\c{a}}k, and  Professor A. Khater for providing excellent working conditions and the financial support.

D. Szcz{\c{e}}{\'s}niak would like to acknowledge the financial support under the "Young Scientists" program, provided by the Dean of the Faculty of Mathematics and Science JDU Professor Z. St{\c{e}}pie{\'n} (grant no. DSM/WMP/1/2011/17).

Some calculations have been conducted on the Cz{\c{e}}stochowa University of Technology cluster, built in the framework of the
PLATON project, no. POIG.02.03.00-00-028/08 - the service of the campus calculations U3. 
\end{acknowledgments}


%
\end{document}